\documentclass[aps,prd,twocolumn,reprint,showpacs,superscriptaddress,nofootinbib,amsfonts,amssymb,amsmath]{revtex4-1}  
\usepackage[english]{babel}
\usepackage[utf8]{inputenc}
\usepackage{graphicx}  
\usepackage{dcolumn}   
\usepackage{bm}        
\usepackage{amssymb}   
\usepackage{setspace}
\usepackage[hidelinks]{hyperref}
\usepackage{verbatim}
\usepackage{indentfirst}
\usepackage{textcomp}
\usepackage{braket}
\usepackage{amsmath}
\usepackage{blindtext}
\usepackage{xcolor}
\usepackage{amsfonts}
\usepackage{natbib} 

\begin{document}
\author{Roberta Chiovoloni}
\affiliation{Department of Physics, University of Swansea, Sketty, Swansea SA2 8PQ, UK}
\author{Giovanni Montani}
\affiliation{Department of Physics, University of Rome “La Sapienza”, P.le Aldo Moro 5, Roma, 00185, Italy}
\affiliation{ENEA C. R. Frascati, Via E. Fermi 45, Frascati, Roma 00044, Italy}
\author{Valerio Cascioli}
\affiliation{Department of Physics, University of Rome “La Sapienza”, P.le Aldo Moro 5, Roma, 00185, Italy}
\affiliation{INFN Roma1, Rome, Italy}

\title{Quantum dynamics of the corner of the Bianchi IX model in the WKB approximation}

\begin{abstract}
In this paper we analyze the Bianchi IX Universe dynamics within the corner region associated to the potential term which the spatial curvature induces in the Minisuperspace.
The study is done in the vacuum and in the presence of a massless scalar field $\phi$ and a cosmological constant term $\Lambda$.
We investigate the dynamics in terms of WKB scenario for which the isotropic Misner variables (the volume) and one of the two anisotropic ones (and $\phi$ when present) are treated on a semi-classical level, while the remaining anisotropy degree of freedom, the one trapped in the corner, is described on a pure quantum level. 
The quantum dynamics always reduces to the one of a time-dependent Schr\"{o}edinger equation for a harmonic potential with a time dependent frequency. 
The vacuum case is treated in the limits of a collapsing and an expanding Universe, while the dynamics in presence of $\phi$ and $\Lambda$ is studied for $t \rightarrow \infty $. 
In both analysis the quantum dynamics of the anisotropy variable 
is associated to a decaying standard deviation of its probability density, corresponding to a suppression of the quantum  anisotropy associated. In the vacuum case, the corner configuration becomes an attractor for the dynamics 
and the evolution resembles that one of a Taub 
cosmology in the limit of a non-singular initial Universe. This suggests that if the Bianchi dynamics enters enough the potential corner
then the initial singularity is removed and a Taub picture emerges. 
The case when $\phi$ is present well mimics the De-Sitter phase of an inflationary Universe. Here we show that both the classical and quantum anisotropies are exponentially suppressed, so that the resulting dynamics corresponds to an isotropic closed Robertson-Walker geometry. 
\end{abstract}

\maketitle

\section{Introduction}
The Bianchi IX model \cite{Belinsky:1970ew, Arnowitt:1960es, Montani:2011zz} has a relevant role in the study of the cosmological dynamics since, despite its spatial homogeneity, it possesses typical features of the generic cosmological solution \cite{Belinsky:1982pk, Kirillov:1993aa, Montani:1999aa}, like a chaotic time evolution of the cosmic scale factors near the cosmological singularity \cite{Imponente:2001fy, Montani:2011zz}.
In the Hamiltonian representation, the Bianchi IX dynamics can be reduced to that one of a two-dimensional point-particle in a time dependent potential \cite{Arnowitt:1960es, Misner:1974qy}. The chaotic features of the model correspond to an infinite sequence of bounces of the point particle against the potential walls, which in the representation based on the so-called Misner-Chitre-like variables can be shown to induce an ergodic evolution, having also a significant degree of stochasticity \cite{Chitre:1972aa, Misner:1974qy, Imponente:2001fy}. 

In the asymptotic limit to the cosmological singularity, the potential term of the Bianchi IX dynamics resembles an infinite well having the morphology of an equilateral triangle. 
However, three open corners appear in the vertices of such a triangular configuration; they correspond to the non-singular Taub cosmology \cite{Misner:1969aa}, which defines the limit when the Bianchi dynamics is associated to two equal scale factors of the three possible independent ones. 

It was shown \cite{Imponente:2001fy} (see also the original literature therein) that, during its evolution toward the initial singularity, there is always a situation where the point-particle is deeply inside one of the corner and the two very close scale factor rapidly oscillate. 
In \cite{Belinsky:1970ew} the authors defined this regime as "small oscillations". 
\\

However, despite the Bianchi IX model spends a long time in the small oscillations configuration, it is a well-known result \cite{Montani:2011zz} that, sooner or later, it escapes this regime  to restore the standard dynamics in the central region of the potential well and, furthermore, the probability that small oscillations take place again is strongly suppressed. 

In the present analysis we study the situation in which the Bianchi IX dynamics is trapped in a corner of the potential, but the small degree of anisotropy, which is oscillating, is in a quantum regime. 

We consider two different cases: Bianchi IX in the vacuum , and in presence of a massless scalar field $\phi$ and a cosmological constant $\Lambda$, able to mimic an inflationary-like paradigm. 

The paradigm we are addressing corresponds to the WKB proposal of Vilenkin \cite{Vilenkin:1988yd} for the interpretation of the wave-function of a small quantum subsystem of the Minisuperspace. 

The idea is that a part of the primordial Universe (the volume, the macroscopic anisotropy and $\phi$ when present) has reached a quasi-classical limit and can therefore play the role of a clock, for the small quantum subsystem, which corresponds to the small anisotropy variable trapped in the corner, near a zero value.

For an implementation of the same scenario in the case of a quasi-isotropic Bianchi IX Universe and a quasi-isotropic Taub Universe see \cite{Battisti:2009qd} and \cite{DeAngelis:2020wjp}, respectively. 

In both these cases, it has been shown that the small quantum degrees of freedom are naturally suppressed by the Universe exponential expansion during the De-Sitter phase. 
For a better characterization of the concept of smallness of the quantum subsystem in the Vilenkin BKL scenario, see \cite{Agostini:2017aa}.

In the vacuum case we distinguish two different situations corresponding to the expanding or collapsing behaviour of the Universe respectively. When the volume expands and the classical anisotropy increases towards larger values, the standard deviation of the probability
distribution associated to the small quantum anisotropy degree of freedom is damped to zero and the Universe asymptotically approaches a Taub cosmological model \cite{Misner:1969aa}.
As a result, if the point-Universe enters sufficiently into the corner, this configuration becomes an attractor and
the quantum anisotropy is increasingly damped.

If we consider this picture in the direction of a collapsing Universe instead, we get that the frequency of the harmonic oscillator
associated to the quantum anisotropy takes a constant value. Therefore, the classical component of the Universe takes the form of a Taub Universe,
possessing a small fluctuating additional anisotropy. It is known \cite{Misner:1969aa} that the Taub model has a singularity in the future, but a non-singular finite Universe volume in the past. Thus if we start with a point Universe entering the corner backward in time,
thought as the past of the considered framework, the approach to the initial singularity would be stopped.

In this respect, differently from the pure classical behaviour (see \cite{Belinsky:1970ew,Montani:2011zz}), in a WKB
scenario a la Vilenkin, in which the small anisotropy is thought as a quantum degree of freedom, the existence of the initial singularity could be removed. 
The backward extension of a Mixmaster dynamics \cite{Arnowitt:1960es} sooner or later would deeply enters the corner and the limiting initial configuration of the Universe would be a finite volume Universe, endowed with a small
stationary distribution for the relic quantum anisotropy.
This conjecture could offer a more general paradigm if we recall that the Bianchi IX model is the prototype for the generic cosmological solution
\cite{Belinsky:1982pk,Montani:2011zz}.

When $\phi$ and $\Lambda$ are included in the dynamics, we consider the limit of an asymptotic exponentially expanding Universe, according to a De-Sitter phase of an inflationary paradigm. We show that both the classical macroscopic anisotropy, and the small quantum one are exponentially suppressed as the volume expands. By other words, we are implementing a new dynamical scheme for the isotropization of the Bianchi IX dynamics. This issue completes the analysis in \cite{Battisti:2009qd}, where the depicted scenario corresponds to the case of two small quantum anisotropies, i.e. the  case when the point-particle is close to the potential center. 

The present study seems to be of more cosmological interests, since we expect, due to the time reversibility of the Einsteinian dynamics, that also in the expanding picture the Bianchi IX Universe spends long time in the corner configuration. This consideration makes plausible that, on one hand the small anisotropy degree of freedom is in a quantum regime, and on the other hand, the cosmological constant term has time to grow, and therefore the de-Sitter phase has time to start. 

The paper is structured as follows: in \eqref{hamilton_form} we introduce the Hamiltonian formulation of the Mixmaster model and derive the point-Universe approximation, in \eqref{section_mix} we describe the quantum behaviour of the model and the Vilenkin approach to the wave function of the Universe, in \eqref{sec_4} we show the implementation of the Vilenkin idea to the Mixmaster and discuss the obtained results and finally, in \eqref{sec_con} we summarize the conclusions.

\section{Hamiltonian Formulation of the Mixmaster model}
\label{hamilton_form}
The importance of the Hamiltonian formulation of the Mixmaster model, obtained following  the ADM method \cite{Montani:2011zz}, relies on the fact that it shows how it is possible to reduce the dynamics of the Bianchi IX model to the dynamic of a two-dimensional point particle performing an infinite series of bounces inside a potential well. 
\\
Here and in the following we are going to adopt $\hbar = c = 1$.
\\

The line element for this model is 
\begin{equation}
ds^2 = N^2(t)dt^2 - e^{q_a}\delta_{ab}\omega^a\omega^b
\end{equation}
where $\omega^i$ are 1-form depending on the Euler angles $\theta \in [0, \pi]$, $\phi \in [0, 2\pi] $, $\psi \in [0, 4\pi]$, and $N$ is the lapse function. 
Considering the Einstein-Hilbert action
\begin{equation}
S = -\frac{1}{2k} \int d^4x \sqrt{-g}R = \int dt (p_a\dot{q}^a-N\mathcal{H}_B)
\end{equation}
where $p_a$ are the conjugate momenta to the generalized coordinates $q^a$, and $k = 8\pi G$, the \textit{Hamiltonian density} can be written as 
\begin{equation}
\label{H_B1}
\mathcal{H}_B = \frac{k}{8\pi^2 \sqrt{\eta}} \left[\sum_a (p_a)^2 - \frac{1}{2}\left(\sum_b p_b\right)^2 - \frac{64 \pi^4}{k^2} \eta ^3R\right]  
\end{equation}
where $\eta = det (\eta_{ab}) = exp [\sum_a q_a]$ and the last term on the right-hand side can be 
viewed as a potential for the dynamics. 
\\

To obtain a Hamiltonian that resembles the one of a point-particle, it is necessary to diagonalize the kinetic part introducing the following variables: 

\begin{equation}
\begin{cases}
	q_1 = 2 (\alpha + \beta_+ +\sqrt{3}\beta_-)\\

	q_2 = 2 (\alpha + \beta_+ -\sqrt{3}\beta_-)\\

	q_3 = 2(\alpha-2\beta_+)\\
\end{cases}
\end{equation}

where $\alpha$, $\beta_{\pm}$ are the \textit{Misner variables}, introduced by Misner in \cite{Arnowitt:1960es}:  $\alpha$ describes the volume of the Universe, $\beta_{\pm}$ describe the anisotropy degrees of freedom.

The introduction of the Misner variables allows us to write the super Hamiltonian constraint 
for the Bianchi IX Universe 
in the ADM formalism \cite{Arnowitt:1960es} as 
\begin{equation}
\label{Hamiltonian constraint}
\begin{split}
\mathcal{H}_{IX} &= \frac{N k}{3(8\pi)^2} e^{-3\alpha} (-p^2_{\alpha} + p^2_+ + p^2_-+ p_{\phi}^2 ) + \\
&+\frac{N k}{3(8\pi)^2} e^{-3\alpha} \left[\frac{3(4\pi)^4}{k^2} e^{4\alpha} V_{IX}(\beta_-, \beta_+) + \Lambda e^{6\alpha}\right] = 0
\end{split}
\end{equation}

where ($p_{\alpha}$, $p_{\pm}$) are the conjugate momenta to ($\alpha$, $\beta_{\pm}$) and we added a classical scalar field $\phi$ and the cosmological constant $\Lambda$ in order to obtain an inflationary scenario. \\
$V_{IX} (\beta_-, \beta_+)$ is the potential of the Bianchi IX model and is given by 
\begin{equation}
\label{potential}
\begin{split}
V_{IX} (\beta_-, \beta_+) = &e^{-8\beta_+}-4e^{-2\beta_+} \cosh (2\sqrt{3}\beta_-) + \\
&+ 2 e^{4\beta_+}  [\cosh (4\sqrt{3}\beta_-)-1].
\end{split}
\end{equation}

\begin{center}
\begin{figure}[h]
\includegraphics[scale=0.32]{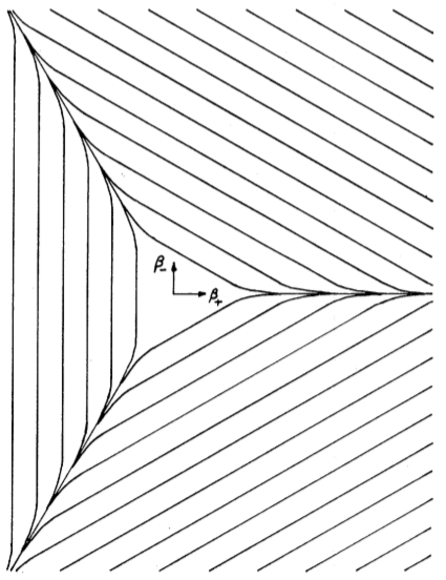}
\caption{Equipotential lines for the Bianchi IX potential \eqref{potential} in the plane $ (\beta_+ , \beta_ -)$. }\label{potentialplot}
\end{figure}
\end{center}

As seen from Fig.\eqref{potentialplot} this function has the symmetry of an equilateral triangle with steep exponential walls and three open angles. 
\\

The expressions for the equipotential lines for large values of $\vert \beta_+ \vert$ and small  $\vert \beta_- \vert $ are: 

\begin{equation}
\label{potential_approx1}
V_{IX} \propto 
	\begin{cases}
	e^{-8\beta_+}, \ \ \ &\beta_+ \rightarrow - \infty, \vert \beta_-\vert \ll 1 \\
	
	48 \beta_-^2 e^{4\beta_+}, \ \ \ &\beta_+ \rightarrow +\infty, \vert \beta_- \vert \ll 1 	
\end{cases}
\end{equation}
while close to the origin, for $\beta_{\pm} = 0$, 
\begin{equation}
\label{potential_approx2}
V_{IX} \propto (\beta_+^2 + \beta_-^2).  
\end{equation}

The Hamiltonian approach provides the following equations of motion 
\begin{equation}
\label{alpha_motion}
\dot{\alpha} = N \frac{\partial \mathcal{H}_{IX}}{\partial p_{\alpha}}, \ \ \ \ \dot{p}_{\alpha} = N \frac{\partial \mathcal{H}_{IX}}{\partial \alpha},
\end{equation}
\begin{equation}
\label{beta_motion}
\dot{\beta_{\pm}} = N \frac{\partial \mathcal{H}_{IX}}{\partial p_{\pm}}, \ \ \ \ \dot{p}_{\pm} = N \frac{\partial \mathcal{H}_{IX}}{\partial \beta_{\pm}}.  
\end{equation}




The Universe evolution at this point is described as the motion of a \textit{point-like particle} governed by \eqref{potential_approx1}
and \eqref{potential_approx2} 
and is characterized by a sequence of bounces against the potential wall, when the system evolves towards the singularity \cite{Montani:2011zz,Cianfrani:2014aa}.

\section{Quantum behaviour of the Mixmaster model}
\label{section_mix}
In this section we are going to briefly introduce the \textit{Wheeler-deWitt equation} (WDW) and to show how it is used in \cite{Vilenkin:1988yd} in order to obtain an interpretation of the wave function of the Universe.\\

The WDW equation describes the quantum behaviour of the Universe and it can be seen as the quantum version of the superhamiltonian constraint $\eqref{Hamiltonian constraint}$. 

To canonically quantize a system, the required commutation relation is 
\begin{equation}
[\widehat{q}_a, \widehat{p}_b] =  i \delta_{ab}
\end{equation}
which is satisfied for $\widehat{p}_a = -i\hbar \partial_a$. 

Therefore, imposing the constraint equation $\eqref{Hamiltonian constraint}$ 
and replacing the canonical variables with their operators in order to select the physically allowed states, we obtain the WDW equation
\begin{equation}
\label{WdW_General}
\widehat{\mathcal{H}}\Psi = (\nabla^2- V)\Psi = 0
\end{equation} 
where $\Psi$ is the 
\textit{wave function of the Universe}, which provides information about the physical state of the Universe. 

It is important to notice that in the general formulation of the WDW equation, $\Psi$ is defined on a \textit{superspace}, intended as the infinite dimensional space of all the possible three-metric where the wave function is defined, while in the present paper it is defined on a \textit{mini-superspace}; this is obtained restricting the number of degrees of freedom of the metric to a finite number by imposing symmetries. 
This simplification is possible since we are focusing on homogeneous models where only three degrees of freedom (the three different scale factors) are allowed. 

With this in mind, we can finally write the WDW equation for the Bianchi IX model 
\begin{equation}
\label{WdW_BianchiIX}
\begin{split}
\widehat{\mathcal{H}}_{IX} \Psi = \frac{N k }{3(8\pi)^2}e^{-3\alpha} [\partial_{\alpha}^2 - \partial^2_{+} - \partial^2_- -\partial_{\phi}^2 ] \Psi + \\
+\frac{N k }{3(8\pi)^2}e^{-3\alpha} \left[\frac{3(4\pi)^4}{k^2}e^{4\alpha}V_{IX}+ \Lambda e^{6\alpha}\right] \Psi = 0
\end{split}
\end{equation}
where $\Psi = \Psi (\alpha, \beta_+, \beta_-, \phi)$. 

\subsection{Interpretation of the wave function of the Universe}

 
In this section we give a probabilistic interpretation of the wave function of the Universe, following \cite{Vilenkin:1988yd}. 

The initial assumption is that it is possible to separate semiclassical variables $h_{\alpha}$ from the quantum ones $q^{\nu}$. Hence \eqref{WdW_General} becomes:
\begin{equation}
(\nabla^2_0 -V_0- H_q) \Psi = 0 
\end{equation}
where $(\nabla^2_0 -V_0) \equiv H_0$ is the part of the WdW obtained neglecting the quantum variables and their momenta. 

One of the hypothesis of the Vilenkin idea is that it is possible to neglect the effect of the quantum variables on the dynamics of the semiclassical ones; therefore he imposed the following condition 
\begin{equation}
\frac{H_q \Psi}{H_0 \Psi} = O (\hbar)
\end{equation} 

Such requirement is physically reasonable since the semiclassical properties of a cosmological model, as well as the smallness of a quantum subsystem, are expectably linked to the fact that the Universe is large enough.

The wave function of the Universe can be written as a WKB expansion 
\begin{equation}
\label{wave_function_of_the_Universe}
\Psi = A(h) e^{i S(h)} \chi(h,q) \equiv \Psi_0 \chi
\end{equation}
where $\Psi_0 $ is such that 
$(\nabla_0^2 - V_0) \Psi_0 =0$ . 

Performing a WKB expansion the equations obtained for the 0-th and 1-st order in $\hbar$ are 
\begin{equation}
\label{H-J}
g^{\alpha\beta} (\nabla_{\alpha} S)(\nabla_{\beta} S) + V = 0
\end{equation}
and 
\begin{equation}
\label{continuity_base}
2\nabla A \nabla S + A \nabla^2 S + 2i \nabla S\nabla\chi - H_q\chi =0.
\end{equation}
The $\eqref{continuity_base}$ can be decoupled in a pair of equations using the \textit{adiabatic approximations}, 
for which $\vert\partial_{q} \chi \vert  \gg \vert \partial_{q} A \vert$
: this states that the semiclassical evolution is principally contained in the semiclassical part of the wave function, while the quantum part depends on it parametrically. 
From $\eqref{continuity_base}$ we obtain 
\begin{equation}
\frac{1}{A}\nabla (A^2 \nabla S) = 0, \ \ \ \ 2i\nabla S \nabla \chi -H_q\chi =0.
\end{equation}

The first equation represents the conservation of the probability current, while the second equation can be rewritten as a Schr\"{o}edinger-like equation
\begin{equation}
i\hbar \frac{\partial \chi}{\partial t }= N H_q \chi.
\end{equation}

Therefore, following the steps in \cite{Vilenkin:1988yd} we can get two different parts of the probability current, one related to the components of the classical subspace and the other one to those in the quantum one
\begin{equation}
\label{probability_distr_Vilenkin}
\rho(h, q, t) = \rho_0(h, t)\vert \chi (q, h(t), t) \vert ^2
\end{equation}
where $\rho_0(h,t)$ and $\vert \chi \vert ^2$ are the probability distribution for the $h_{\alpha}$ and the $q^{\nu}$ quantum variables. In particular, $\rho_0(h,t) = \vert A (h) \vert^2$ 

\section{Application of the Vilenkin approach to Bianchi IX}
\label{sec_4}
To describe Bianchi IX 's dynamics close to the singularity we use the Misner variables $\alpha$, $\beta_+$ and $\beta_-$ and we separate them into semi-classical and quantum, in order to use the Vilenkin approach; we choose $\alpha$ and $\beta_+$ to be semi-classical variables, while $\beta_-$ is the quantum one.

Therefore, looking at Fig.\eqref{potentialplot}, we are considering as initial condition for the point-Universe the right corner of the potential, where $\beta_+ \rightarrow +\infty$ and $\vert \beta_- \vert \ll 1.$ \\
In the following analysis we will  include a massless scalar field $\phi$, for which  $\dot{\phi} \ll V(\phi)$, and preferring to work in a \textit{synchronous frame} we set $N(t)=1$. \\
This is easily explained looking at the relation between  the synchronous time and the lapse function : 
\begin{equation}
    t_s = \int N(t') dt'.
\end{equation}

Recalling \eqref{wave_function_of_the_Universe}, the wave function of the Universe in the Mixmaster model  can be written in its explicit form as 
\begin{equation}
\label{wave_funct_universe}
    \Psi = \Psi_0 \chi = A(\alpha, \beta_+, \phi) e^{i S(\alpha, \beta_+, \phi)} \chi(\alpha, \beta_+, \phi, \beta_-).
\end{equation}

Substituting this expression into \eqref{continuity_base} and using the \textit{adiabatic approximation} we obtain 
\begin{equation}
\label{current conservation }
    2\left[\frac{\partial A}{\partial \alpha} \frac{\partial S}{\partial \alpha} - \frac{\partial A}{\partial \beta_+} \frac{\partial S}{\partial \beta_+} - \frac{\partial A}{\partial \phi} \frac{\partial S}{\partial \phi}\right] + A\left[\frac{\partial^2 S}{\partial \alpha^2} - \frac{\partial^2S}{\partial \beta_+^2} - \frac{\partial^2 S}{\partial \phi^2}\right] = 0
\end{equation}
\begin{equation}
\label{schroedinger}
    2i\left[\frac{\partial S}{\partial \alpha}\frac{\partial \chi}{\partial \alpha} - \frac{\partial S}{\partial \beta_+}\frac{\partial \chi}{\partial \beta_+}\right] = \mathcal{H}_q \chi
\end{equation}

where $\mathcal{H}_q  = -\partial_-^2 + 16 e^{4(\alpha + \beta_+)} \beta_-^2$ given the initial condition chosen for $\beta_+$ and $\beta_-$. 

Please note that from now on we use $\mathcal{V}$ to indicate the approximate expression $V_{IX} \propto 48 \beta_-^2 e^{4\beta_+}$. 

Eq \eqref{current conservation } can be viewed as a continuity equation, while \eqref{schroedinger} can be rewritten as a Schr\"{o}edinger-like equation using the equation of motions \eqref{alpha_motion} and \eqref{beta_motion} and introducing a new time variable $\tau$: 
\begin{equation}
\label{new_time_variable}
    \tau = c \int e^{-3\alpha} dt 
\end{equation}
where $c$ is only a numerical factor. 
Following this simplification, \eqref{schroedinger} becomes 
\begin{equation}
\label{to_solve}
    i \frac{d \chi}{d \tau} = \mathcal{H}_q \chi. 
\end{equation}

\subsection{Resolution of the Schr\"{o}edinger equation}
The quantum probability distribution for the wave function of the universe, as can been seen in \eqref{probability_distr_Vilenkin}, is given by $\vert \chi \vert^2$, therefore for its computation it is necessary to solve \eqref{to_solve}. 

Substituting $\mathcal{H}_q$ explicitly with its expression, equation \eqref{to_solve} becomes
\begin{equation}
\label{explicit_Schr}
    i\frac{d\chi }{d \tau} = \left(p_-^2 +  16 e^{4(\alpha + \beta_+)}\beta_-^2\right)\chi
\end{equation}
which can be viewed as the Schr\"{o}edinger equation of a harmonic oscillator with time-dependent frequency and unitary mass if we impose $\omega^2(\tau) \equiv 16 e^{4(\alpha + \beta_+)}$ and redefine the time variable $\tau ' = 2\tau$. 
Note that in the following we will use $\tau$ instead than $\tau'$. 

In \cite{Lewis:1969a,Lewis:1967b,Lewis:1968yx} the authors developed a method to obtain
eigenvectors and eigenvalues for this particular Schr\"odinger equation using the invariant method. As summarized in \cite{Pedrosa:1997aa} the general solution of an equation of the form \eqref{explicit_Schr} is given by:

\begin{equation}
\label{general_solution}
\chi = \sum_n c_n e^{i\alpha_n(\tau)}\phi_n (\beta_-, \tau) = \sum_n c_n \chi_n(\beta_-, \tau)
\end{equation}
where $c_n$ are numerical coefficients that weight the different $\chi_n$, 
\begin{equation}
\alpha_n (\tau) = -(n+\frac{1}{2})\int_0^\tau \frac{1}{\rho^2} d\tau' 
\end{equation}
\begin{equation}
\chi_n(\beta_-, \tau) = \Omega _n \exp \left[\frac{i}{2\hbar}\left(\frac{\dot{\rho}}{\rho}+\frac{i}{\rho^2}\right)\beta_-^2 ] \mathcal{H}_n(\frac{\beta_-}{\rho})\right]
\end{equation}
where $\Omega_n = \left[\frac{1}{(\pi )^{1/2} n! 2^n \rho}\right]^{1/2}$ and $\rho$ is a \textit{c}-number quantity satisfying 
\begin{equation}
\label{rhot}
\rho'' +\omega^2(\tau)\rho-\frac{1}{\rho^3} = 0
\end{equation}
where the $'$ indicates a differentiation respect to the time variable $\rho$ depends to, which is $\tau$. 

It is usually complicated to solve \eqref{rhot} analytically, but in \cite{Lewis:1967b} the authors developed a method that allows us to have the explicit expression of the $\rho$ as a linear combination of $f(\tau)$ and $g(\tau)$, linear solutions of 
\begin{equation}
\label{rho_2}
    \frac{\partial^2 q}{\partial \tau ^2} + \omega^2(\tau)q = 0  .
\end{equation} 

\subsection{Bianchi IX in the vacuum}
\label{vacuum}
As first step, we study the dynamical evolution of the Mixmaster model in the simplest case: the vacuum.
In this case the Hamiltonian is simply (starting from \eqref{Hamiltonian constraint}) 
\begin{equation}
    \mathcal{H} = e^{-3\alpha} K (-p_{\alpha} ^2 + p_+^2 + p_-^2 + \mathcal{V})
\end{equation}
where K is a numerical coefficient and the quantum part of the Hamiltonian $\mathcal{H}_q$ is given by 
\begin{equation}
\mathcal{H}_q = -p^2_- + \omega^2(\tau) = \left(p_-^2 +  16 e^{4(\alpha + \beta_+)}\beta_-^2\right) 
\end{equation}
as already stated in \eqref{explicit_Schr}. 

To compute the solution of \eqref{explicit_Schr} we need to write $\omega^2(\tau)$ explicitly as a function of the time variable and this can be done starting from \eqref{alpha_motion}, \eqref{beta_motion} and \eqref{Hamiltonian constraint} . 
In particular
\begin{equation}
\label{alpha_equation}
    \dot{\alpha} = \frac{\partial \alpha}{\partial t} = \frac{\partial \mathcal{H}_0}{\partial p_{\alpha}} = - 2 p_{\alpha} k e^{-3\alpha}
\end{equation}
\begin{equation}
\label{momentum_constrain}
\dot{p}_{\alpha} = \frac{\partial p_{\alpha}}{\partial t} = -\frac{\partial \mathcal{H}_0}{\partial \alpha}= -3 \mathcal{H} = 0
\end{equation}
where $\mathcal{H}_0 = e^{-3 \alpha} k (-p_{\alpha}^2 + p_+^2) $. \\

Integrating \eqref{alpha_equation} through separation of variables, and using the result of \eqref{momentum_constrain}, which states that $p_{\alpha}$ has a constant value, we get 
\begin{equation}
e^{3 \alpha} = 6 \vert p_{\alpha}\vert  K t 
\end{equation}
which gives 
\begin{equation}
\label{alpha_t}
\alpha(t) = \frac{1}{3} \log{6 \vert p_\alpha \vert } K + \frac{1}{3}\log{t} 
\end{equation}

It is worth noticing that, in the calculation above we used the absolute value of $p_{\alpha}$; looking at \eqref{alpha_equation} we see that $\dot{\alpha}$, which denotes how the volume of the Universe changes with time, has the opposite sign of $p_{\alpha}$. Since our study is based on an expanding Universe and therefore we need $\dot{\alpha} >0$, we impose $p_{\alpha} <0$. \\
Using \eqref{alpha_t} in \eqref{new_time_variable} we obtain
\begin{equation}
\label{tau_vacuum}
    \tau(t) = \frac{1}{6\vert p_{\alpha} \vert } \log{t}
\end{equation}
that can be substitute it in \eqref{alpha_t} to give
\begin{equation}
\label{alpha_vacuum}
\alpha (\tau) = \frac{1}{3} \log{6 \vert p_{\alpha} \vert K} + 2 \vert p_{\alpha} \vert \tau. 
\end{equation}

Given \eqref{tau_vacuum}, and the asymptotic behaviour of the synchronous time variable  $t$, $0 < t < \infty$, we have that $ -\infty < \tau < \infty $. 

As mentioned in the introduction, for the study of the vacuum , we distinguished two different situations, which correspond to the expanding and collapsing behaviour of the Universe. 

\subsubsection{Bianchi IX in the vacuum: expanding Universe}

In this section we consider the dynamical evolution when the semi-classical anisotropy variable $\beta_+$ increases towards larger values in the direction of an expanding Universe, that means $\dot{\beta}_+(t) > 0$. \\
To find the explicit expression for $\beta_+(t)$ and consequently for $\beta_+(\tau)$, we follow the same arguments presented in the computation of $\alpha(t)$. 

In particular from \eqref{beta_motion} 
\begin{align}
\label{beta_+_t}
\dot{\beta}_+(t) &= \frac{\partial \mathcal{H}_0}{\partial p_+} = 2 p_+ K e^{-3\alpha} \notag \\
\beta_+(t) &= \frac{1}{3} \frac{p_+}{\vert p_{\alpha} \vert} \log{t} + \beta_0
\end{align}

\begin{equation}
    \dot{p_+} = -\frac{\partial \mathcal{H}_0 }{\partial \beta_+} = 0  \ \rightarrow \ p_+ = const. 
\end{equation}

The ratio $p_+ / \vert p_{\alpha} \vert$ can be simplified using the Hamilton-Jacobi equation 
\begin{equation}
\label{HJ}
\left( \frac{\partial S }{\partial \alpha} \right) ^2 - \left( \frac{\partial S }{\partial \beta_+} \right) ^2 = 0 \rightarrow p_{\alpha} = \pm p_+ 
\end{equation}
and taking into account that we are interested in studying what happens for $ t \rightarrow \infty$ (hence $\dot{\alpha} > 0)$ and deeply inside the corner, therefore for $\dot{\beta_+} > 0$: this translates in the condition $p_+ > 0 $. 

Substituting \eqref{alpha_vacuum} and \eqref{tau_vacuum} into \eqref{beta_+_t} we obtain 
\begin{equation}
\beta _{+} (\tau) = \beta_0 + 2 \vert p_{\alpha} \vert \tau 
\end{equation}

The frequency of the harmonic oscillator becomes 
\begin{align}
    \omega^2(\tau) &= 16 e^{4(\alpha(\tau) + \beta_+(\tau))} \notag \\ 
    &= 16 e^{4(\alpha_0+\beta_0)}e^{16\vert p_{\alpha} \vert \tau}  \propto C e^{k \tau},
\end{align}
with $k$ and $C$ constants. 
\\
The solution of equation \eqref{explicit_Schr} can be obtained solving \eqref{rho_2} to find $\rho(\tau)$. 
The 2 independent solutions are : 
\begin{equation}
    f(\tau) = J_0 \left[\frac{2\sqrt{C}\sqrt{e^{k\tau}}}{k}\right]
\end{equation}
\begin{equation}
    g(\tau) = N_0\left[\frac{2\sqrt{C}\sqrt{e^{k\tau}}}{k}\right]
\end{equation}
where $J_0$ and $N_0$ represent the Bessel functions of the
first and the second kind. 

Combining them together we obtain
\begin{widetext}
\begin{equation}
\label{rho_vuoto}
\rho(\tau) = \frac{\pi}{2 k}  \sqrt{J_0^2\left[\frac{2\sqrt{C}\sqrt{e^{k\tau}}}{k}\right] + \frac{64 k^2 N_0^2\left[\frac{2\sqrt{C}\sqrt{e^{k\tau}}}{k}\right] }{\pi^2} + \frac{8\sqrt{3} k J_0 \left[\frac{2\sqrt{C}\sqrt{e^{k\tau}}}{k}\right] N_0 \left[\frac{2\sqrt{C}\sqrt{e^{k\tau}}}{k}\right] }{\pi}}. 
\end{equation}
\end{widetext}

\begin{figure}[h]
\includegraphics[scale=0.5]{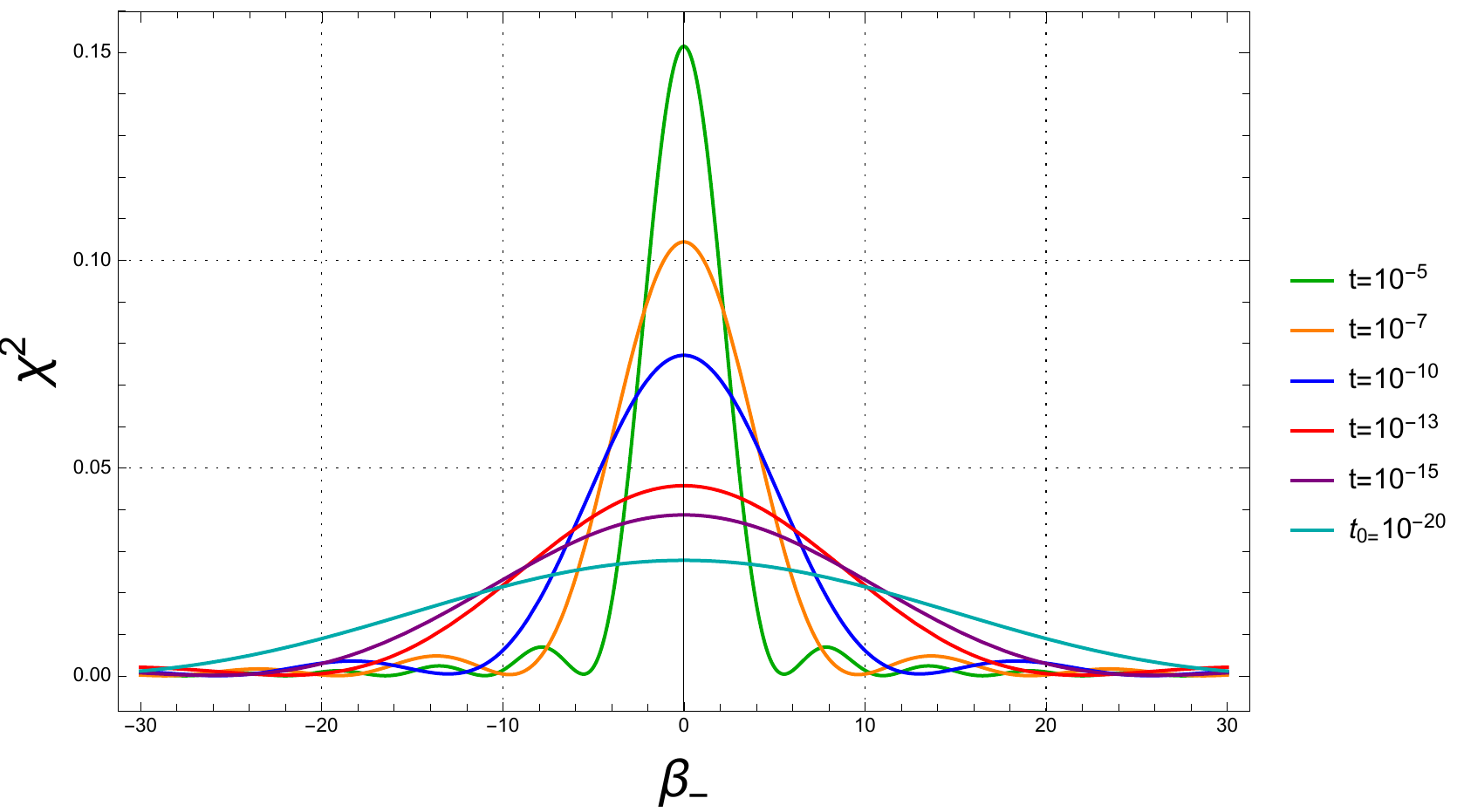}
\caption{Probability density function for different values of the synchronous time variable for Bianchi IX in the  vacuum in the  case $\beta_+ = 2\vert p_{\alpha} \vert \tau$, for an expanding Universe.}\label{fig1}
\end{figure}

Substituting \eqref{rho_vuoto} into \eqref{general_solution} and defining its conjugate, we can finally compute numerically the probability distribution for the quantum
part of the Universe wave function, namely 
\begin{equation}
\label{chi_squared_vacuum}
\vert \chi \left( \tau, \beta _- \right) \vert ^2 = \sum _{n} c_n \chi _n \left( \sum_{m} c_m \chi _m \right)^{*}
\end{equation}
The coefficients $c_n$ are given by 
\begin{equation}
    c_n = \int \chi_0 \chi_n ^* d\beta_-
\end{equation}
where $\chi_0 = \chi_n (\tau_0)$ and it has been chosen such that $\vert \chi_0 \vert ^2$ has a gaussian shape peaked around $\beta_- = 0$. 
\\
We plot $\vert \chi \vert ^2$ as a function of the quantum anisotropic variable $\beta_-$  for different times $t$ in Fig.\ref{fig1}. 

To conclude the study of the probability density of the wave function of the Universe, we calculated the probability density of the semiclassical variables, $\vert A(\alpha, \beta_+) \vert^2$, by variable separation.
 In particular, $A (\alpha, \beta_+)=A_1(\alpha) A_2(\beta_+) $ and results: 
\begin{equation}
A (\alpha, \beta_+) = e^{\frac{W}{p_{\alpha}} (\alpha + \beta_+)}
\end{equation}
where $W$ is a constant.

\subsubsection{Bianchi IX in the vacuum: collapsing Universe }

In this section we consider the dynamical evolution when $\beta_+$ decreases for $t\rightarrow \infty $, since we are interested in changes in the $\vert \chi \vert^2 $ in the direction of a collapsing Universe. \\
The equations of motion and the Hamilton-Jacobi equation do not change compared to those of the previous case \eqref{beta_+_t} and \eqref{HJ}. 

However, the initial assumption in this case is that $\dot{\beta}_+ < 0$, which translates in $p_+ <0$. 
Therefore the semi-classical anisotropic variable becomes:
\begin{equation}
\beta_+(\tau) = \beta_0 -2\vert p_{\alpha} \vert \tau
\end{equation}
and the frequency $\omega^2(\tau)$ reads as
\begin{equation}
\omega^2(\tau) = 16 e^{4(\alpha(\tau) + \beta_+ (\tau))} = 16 e^{4\beta_0}
\end{equation}
which is a constant. 

The solution of \eqref{rhot} can be computed as in \cite{Lewis:1968yx} and it results: 
\begin{equation}
\rho(\tau) = \frac{1}{\sqrt{\omega(\tau)}} = \frac{e^{-\beta_0}}{2}. 
\end{equation}
The eigenfunctions $\chi_n$, which depend on time through $\rho(\tau)$, are constant as well; hence the probability density distribution $\vert \chi \vert^2$ is defined simply by choosing its shape at the initial time. 
Therefore, as the point-Universe moves towards the time singularity, it moves deeply inside the corner ($\dot{\beta}_+ < 0 $), while the probability density $\vert \chi \vert^2$ remains constant. 
\\
In this case, given that $p_{\alpha} = p_{+}$
\begin{equation}
A (\alpha, \beta_+) = e^{\frac{W}{p_{\alpha}} (\alpha - \beta_+)}
\end{equation}
where $W$, as before, is a constant.

\subsection{Bianchi IX model in presence of cosmological constant and scalar field}
\label{generale}
\begin{figure}[h!]
\includegraphics[scale=0.38]{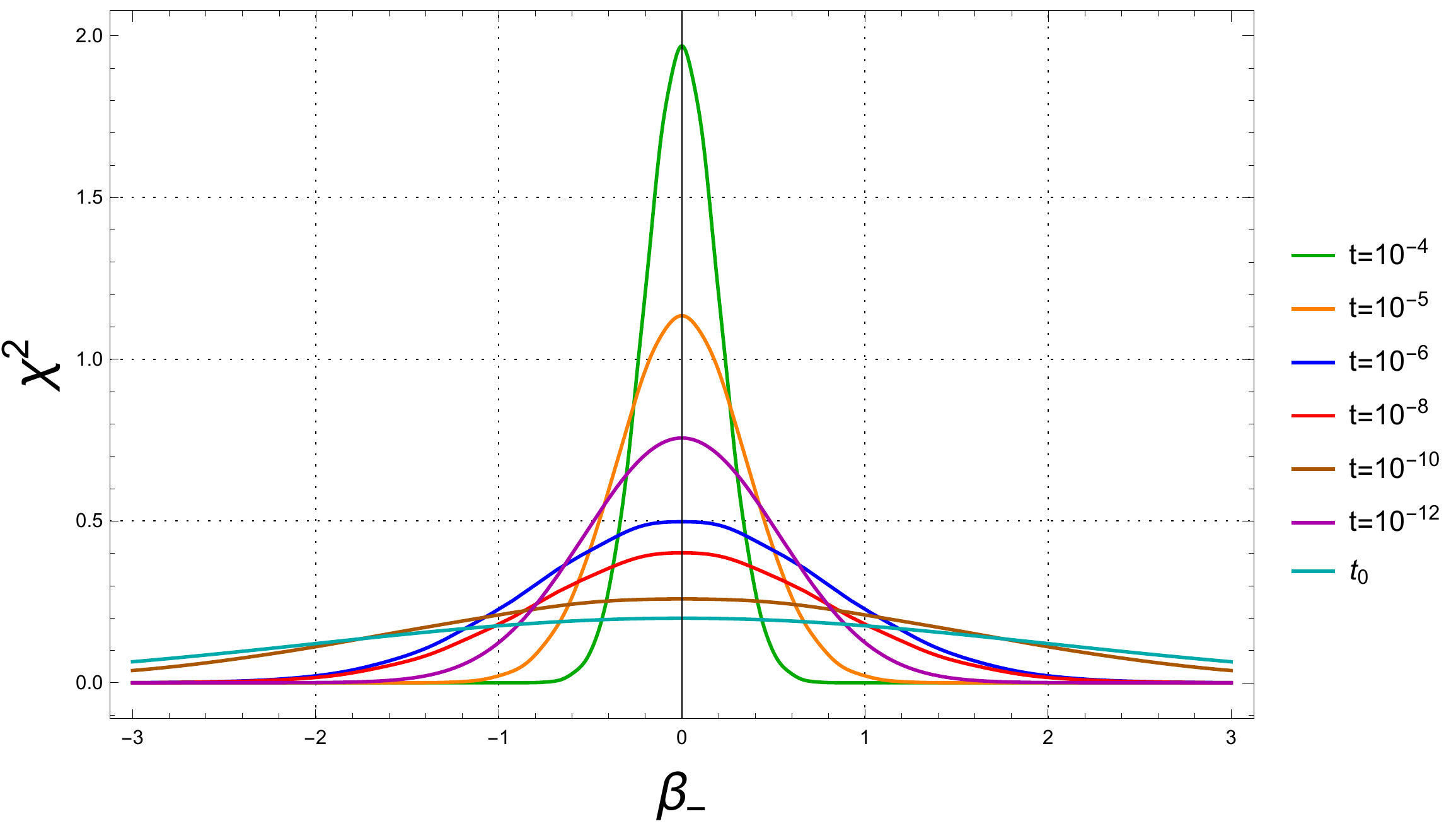}
\caption{Probability density of the quantum subsystem for different values of the time variable in the case of an expanding volume.}
\label{fig2}
\end{figure}

Finally we study the Bianchi IX model in the presence of a cosmological constant $\Lambda$ and a scalar field $\phi$, in order to mimic the inflationary scenario.
In this case the Hamiltonian takes the form
\begin{equation}
\mathcal{H}= e^{-3\alpha}K(-p^2 _{\alpha} + p^2_{+} +p^2 _{-} + p^2 _{\phi} + \mathcal{V} + \Lambda e^{6\alpha} )
\end{equation}
and while $\mathcal{H}_q$ does not change respect to the previous case,
$\mathcal{H}_0 = e^{-3\alpha} K (-p^2_{\alpha} +p_+^2 + p_{\phi}^2+ \Lambda e^{6\alpha})$. 
\\
In this case equations \eqref{H-J} and \eqref{continuity_base} give the following probability distribution for the semi-classical component of the wavefunction of the Universe:
\begin{equation*}
A \left(\alpha, \beta_+, \phi \right) = A_1(\alpha)A_2(\beta_+)A_3(\phi)
\end{equation*}
where
\begin{equation*}
A_1(\alpha) = \frac{Exp\left[\frac{C_1 \tanh^{-1}{\frac{\sqrt{p_+^2 + p_{\phi}^2 + \Lambda e^{6\alpha}}}{\sqrt{p_{+}^2+ p_{\phi}^2}}}}{6 \sqrt{p_{+}^2 + p_{\phi}^2}}\right]}{(p_+^2 + p_{\phi}^2 + \Lambda  e^{6\alpha})^{1/4}} 
\end{equation*}	

\begin{equation*}
A_2(\beta_+) = Exp\left[\frac{C_2}{2p_+} \beta_+\right]
\end{equation*}

\begin{equation}
A_3(\phi) = Exp\left[-\frac{C_1 + C_2}{2 p_{\phi}}\phi\right]
\end{equation}
with $C_1$ and $C_2$ constants. 
Following the same steps of the previous subsection \eqref{vacuum} we can write the full expression for $\tau(t)$, $\alpha(\tau)$ and $\beta_+(\tau)$
\begin{equation}
\label{setH}
\begin{split}
&\tau (t)= \frac{1}{6\sqrt{p^2 _{+} + p^2 _{\phi}}} \log \left\lbrace \tanh \left[\frac{1}{2} \left(6Kt\sqrt{\Lambda} + J \right) \right] \right\rbrace \\
&\alpha (\tau)\! =\! \frac{1}{3}\! \log \left\lbrace \frac{\sqrt{(p^2 _{+} + p^2 _{\phi})}}{\sqrt{\Lambda}} \sinh\left[ 2 \tanh ^{-1} \left(e^{6\tau \sqrt{p^2 _{+} + p^2 _{\phi}}} \right) \right] \right\rbrace \\
&\beta_+ (\tau) = \beta _0 + 2  p_{+} \tau
\end{split}
\end{equation}
and their asymptotic behaviour 
\begin{align}
\label{system_cosmological_constant}
-\infty & < \tau < 0 \notag\\
-\infty & < \alpha < \infty \notag\\
-\infty & < \beta_+ < \beta_0 . s
\end{align}
Given the complexity of the analytic expression of the time and the Misner variables, it was not possible to solve \eqref{rhot} analytically, therefore we computed $\vert \chi \vert^2 $ numerically for different values of $t$.  
\\
The plots obtained are shown in Fig \eqref{fig2}.  \\

It is worth mentioning that, we realized different plots changing the numerical values of $\Lambda$ and the standard deviation $\sigma$ of $\vert \chi_0 \vert^2$. The results did not change respect to those proposed in Fig.\eqref{fig2}. 

\subsection{Discussion of the results}
\label{discussion}
Here we are going to discuss the results obtained above. It is worth reminding that in both cases the initial conditions were such that $ \beta_+ \gg 1 1 $ and $\beta_- \ll 1 $. 


Firstly we are going to analyse the subsection \eqref{vacuum}.
Here we found that, when we evaluate the dynamics of a Gaussian packet initially peaked in $\beta_- =0$, namely $\chi_n(\tau_0)$, as $\beta _+ $ grows in time (meaning that the Universe moves deeper in the corner), the packet tends to peak even more around the value $\beta _- = 0$. 

This can be observed in Fig.\eqref{fig1}. \\
As a consequence, in the proposed representation the corner becomes an attractor for the global system dynamics, and the Universe approaches on a
very good level a Taub cosmological model.
\\
On the contrary, when the point-Universe enters the corner in a backward picture in time, so that the volume is collapsing, the standard deviation of the small quantum anisotropy remains constant, as a consequence of the constant character of the harmonic oscillator frequency. In this case the backward evolution of the Universe would correspond to a Taub Universe, which is no longer a singular cosmology in the past, endowed with a small fluctuating anisotropy degree of freedom, in addition to the macroscopical classical one.\\
Finally we analyse the results of section \eqref{generale} summarized in Fig \eqref{fig2}. \\
In this case we can conclude that there is not a precise trend of the probability density evolution in time, but as the Universe evolves in time, the variable $\beta _+ $ leans to a constant value and the variable $\beta _-$ tends to peak around the value $\beta_- = 0$; thus the presence of the cosmological constant tends to isotropize the Universe.

\section{Concluding remarks}
\label{sec_con}
In this paper we analysed the Bianchi IX cosmology in vacuum, as well as when a massless scalar field $\phi$ and a cosmological constant term $\Lambda$ are present and we limited our attention to the situation in which the point-Universe is trapped in a corner of the scalar curvature potential. \\
The adopted dynamical scheme corresponded to deal with a WKB decoupling of the quasi-classical degrees of freedom, the Misner variables $\alpha$ and $\beta_+$ and $\phi$ when present, from a microscopic fully quantum degree of freedom, the small anisotropy variable $\beta_-$, trapped in the corner. 

In both cases, we had to solve a time dependent Schr\"{o}edinger equation with a quadratic potential, which resembled the equation of a harmonic oscillator with time-dependent frequency.  \\
We demonstrated that, both with and without matter,  the solution of this equation suggests that the small quantum anisotropy $\beta_-$ is strongly suppressed via the dynamics of the quasi-classical variables, . 

In the vacuum case we observe that, if we consider the situation when the point-Universe enters the corner with
an expanding Universe, we find a suppression of the the quantum variable $\beta _-$, as its standard deviation decays in time. 
We conclude that the corner of the potential is an attractor for the point-particle Universe; once the Universe enters, it cannot escape any more. 

Following this analysis, we also studied the limit in which the system approaches the cosmological singularity, where the constant character of the variance associated to the anisotropic variable has a very deep meaning on the whole structure of the Bianchi IX dynamics. 

When $\beta_- \simeq 0$, the resulting cosmology is indistinguishable from a Taub Universe, which is not a singular model in the limit $\alpha \rightarrow - \infty $. 
Since the emergence of a long regime of the classical Bianchi IX dynamics within the a corner has been convincing established \cite {Montani:2011zz}, if the proposed picture is applicable, i.e. the smallness of the $\beta_-$ values justify its quantum treatment, then the singular behaviour of the Bianchi IX Universe could be removed. 

This result, in view of the prototype character of the Bianchi IX cosmology versus the generic cosmological solution \cite{Belinsky:1982pk, Imponente:2001fy}, could have a deep implication on the notion of the cosmological singularity as a general property of the Einstein equations, under cosmological hypotheses.

Finally in the last section \eqref{generale} the study of the Bianchi IX dynamics, performed in the presence of $\phi$ and $\Lambda$, is  developed in expanding picture, i.e. for $\alpha \rightarrow \infty$. 
The aim of this analysis was to mimic the behaviour of the Bianchi IX Universe if the de-Sitter phase, which is associated to the inflationary paradigm for the primordial Universe, takes place when a corner evolution is performed by the point-Universe. 
\\
In this case we have shown that, in the limit of applicability of the WKB proposed scheme, the Universe naturally isotropizes since the classical anisotropy degree of freedom $\beta_+$ is suppressed 
while the fully quantum variable, i.e. $\beta _-$, is characterized by a decaying standard deviation. 
In  other words, if we start with a Gaussian distribution for $\beta_-$, its natural evolution in the future is towards a Dirac delta-function around the zero value. Thus, this study offers a new paradigm for the Bianchi IX cosmology isotropization, based on the idea that the de-Sitter phase is associated with a corner regime of the model.

To conclude, this study generalizes and completes the results discussed in \cite{Battisti:2009qd}, where the Bianchi IX isotropization is faced in the same WKB scenario, but starting with two very small quantum anisotropy variables, i.e. assuming that the de-Sitter phase starts when the point-Universe is in the center of the potential, already near to an isotropic configuration.

\bibliography{refs}
\bibliographystyle{ieeetr}

\end{document}